%% file: Btaunu_LP05_new.tex
\documentclass[aps,prl,preprint,tightenlines,superscriptaddress,showpacs,byrevtex]{revtex4}

\usepackage{graphics}
\usepackage{dcolumn}  
\usepackage{color}
\usepackage{epsf}

\graphicspath{{ps}}

\def\T3M{\tau \rightarrow 3\mu}

\def\E3P{\eta\rightarrow \pi^+\pi^-\pi^0}

\def \Btaunu {B^{-}\rightarrow\tau^{-}\overline{\nu}}
\def \BKnunu {B^{-}\rightarrow K^{-}\nu\overline{\nu}}
\newcommand{\lw}[1]{\smash{\lower1.7ex\hbox{#1}}}
\newcommand{\lww}[1]{\smash{\lower6.7ex\hbox{#1}}}

\begin{document}

\preprint{\vbox{ \hbox{   }
                 \hbox{BELLE-CONF-0566}
                 \hbox{LP2005-201}
                 \hbox{EPS05-541} 
}}

\title{ \quad\\[0.5cm] 
       Search for $B \to \tau \nu$ and $B \to K \nu \bar{\nu}$ Decays with a Fully Reconstructed $B$ at Belle }

\input{author-conf2005.tex}
\noaffiliation

\begin{abstract}

We present a search for the decays $\Btaunu$ and $\BKnunu$ in a $253~\textrm{fb}^{-1}$
data sample collected at the $\Upsilon(4S)$ resonance with the Belle detector 
at the KEKB asymmetric energy $B$ factory. 
Combinatorial and continuum backgrounds are suppressed by selecting a sample
of events with one fully reconstructed $B$. 
The decay products of the $B$ on the other side of the event
are analyzed to search for $\Btaunu$ and $\BKnunu$ decays.
We find no significant evidence for a signal and set 90\% confidence level
upper limits of 
${\cal B}(B^{-}\rightarrow\tau^{-}\overline{\nu}) < 1.8\times 10^{-4}$
and ${\cal B}(B^{-}\rightarrow K^{-}\nu\overline{\nu}) < 3.6\times 10^{-5}$.
All results are preliminary.

\end{abstract}

\maketitle

\tighten

\setcounter{footnote}{0}

The purely leptonic decay $B^{-}\rightarrow\ell^{-}\overline{\nu}$ 
(charge conjugate states are implied throughout the paper)
is of particular interest
since it provides direct measurement of the product of 
the Cabibbo-Kobayashi-Maskawa(CKM) matrix
element $V_{ub}$ and the $B$ meson decay constant $f_{B}$.
In the Standard Model(SM), the branching fraction of the decay 
$B^{-}\rightarrow\ell^{-}\overline{\nu}$ is given as
\begin{equation}
 \label{eq:BR_B_taunu}
{\cal B}(B^{-}\rightarrow\ell^{-}\overline{\nu}) = \frac{G_{F}^{2}m_{B}m_{\ell}^{2}}{8\pi}\left(1-\frac{m_{\ell}^{2}}{m_{B}^{2}}\right)^{2}f_{B}^{2}|V_{ub}|^{2}\tau_{B}
\end{equation}
where $G_{F}$ is the Fermi coupling constant, $m_{\ell}$ and $m_{B}$ are
the charged lepton and $B$ meson masses, $\tau_{B}$ is the $B^{-}$ lifetime.
The dependence on the lepton mass arises from helicity conservation, which 
suppresses the muon and electron channels.
The CKMfitter predicts the $B^{-}\rightarrow \tau^{-}\bar{\nu}$ branching fraction
to be $(9.3 ^{\,+      3.4}_{\,-      2.3}) \times10^{-5}$\cite{Charles:2004jd}.
No evidence for an enhancement relative to the SM prediction
was observed in previous experimental studies.
The most stringent upper limit has been achieved by the BABAR Collaboration : 
${\cal B}(\Btaunu) < 4.2\times 10^{-4}$ at $90\%$ confidence level (C.L.)
\cite{Aubert:2004kz}.

Flavor-changing neutral-current transition such as $b\rightarrow s\nu\bar{\nu}$
occurs in the SM via one-loop box or electroweak penguin diagrams with heavy particles in the loops. Because heavy non-SM particles could contribute additional loop diagrams, 
various new physics scenarios can lead to significant enhancements in the observed rates
\cite{Grossman:1995gt,Bird:2004ts}.
The SM $B^{-}\rightarrow K^{-}\nu\bar{\nu}$ branching fraction has been estimated
to be $(3.8_{-0.6}^{+1.2})\times 10^{-6}$\cite{Faessler:2002ut,Buchalla:2000sk}, 
while the most stringent published
experimental limit is ${\cal B}(\BKnunu) < 5.2\times 10^{-5}$ at $90\%$ C.L.
\cite{Aubert:2004ws}

We use a $253~\textrm{fb}^{-1}$ data sample containing 
$275\times 10^{6}$ $B$ meson pairs collected with the Belle detector
at the KEKB asymmetric energy $e^{+}e^{-}$ ($3.5$ on $8$ GeV) collider
\cite{Kurokawa:2003} operating at the $\Upsilon(4S)$ resonance 
($\sqrt{s} = 10.58$ GeV). 
The Belle detector is a large-solid-angle
magnetic spectrometer consisting of a three-layer silicon vertex detector,
a $50$-layer central drift chamber (CDC), a system of aerogel threshold
$\check{\textrm{C}}$erenkov counters (ACC), time-of-flight scintillation 
counters (TOF), and an electromagnetic calorimeter comprised of
CsI(Tl) crystals (ECL)  
located inside a superconducting solenoid coil that provides a $1.5$ T 
magnetic field. An iron flux-return located outside of the coil is 
instrumented to identify $K_{L}^{0}$ and muons. 
The detector is described in detail elsewhere \cite{belle_detector:2003}.

The strategy adopted for this analysis is to reconstruct exclusively
the decay of one of the $B$ mesons in the event and compare
properties of the remaining particle(s) in the event (referred to as the
signal side) to those expected for signal and background.
All the tracks and photon candidates in the event not used to reconstruct
the $B$ are studied to search for 
$B^{-}\rightarrow\tau^{-}\overline{\nu}$ and $B^{-}\rightarrow K^{-}\nu\overline{\nu}$.
The advantage of having a sample of fully reconstructed $B$ meson is to 
provide a strong suppression of the combinatorial and continuum background
events. The disadvantage is the low efficiency of full $B$ meson 
reconstruction (approximately $0.3\%$).

Fully reconstructed $B$ mesons, $B_{\rm rec}$, are observed in the following decay modes:  
$B^{+}\rightarrow\overline{D}^{(*)0}\pi^{+}$, $\overline{D}^{(*)0}\rho^{+}$, 
$\overline{D}^{(*)0}a_{1}^{+}$ and $\overline{D}^{(*)0}D_{s}^{(*)+}$ 
where $\rho^{+}$ is reconstructed in $\pi^{+}\pi^{0}$ mode ($|M_{\pi^{+}\pi^{0}}-M_{\rho^{+}}| < 0.3~\mbox{GeV}/c^{2}$) and $a_{1}^{+}$ is reconstructed as $a_{1}^{+}\rightarrow \rho^{0}\pi^{+}$ ($|M_{\rho^{0}\pi^{+}}-M_{a_{1}^{+}}| < 0.25~\mbox{GeV}/c^{2}$).
$\overline{D}^{0}$ candidates are reconstructed as 
$\overline{D}^{0}\rightarrow K^{+}\pi^{-}$, $K^{+}\pi^{-}\pi^{0}$,
$K^{+}\pi^{-}\pi^{+}\pi^{-}$, $K_{S}^{0}\pi^{0}$, $K_{S}^{0}\pi^{-}\pi^{+}$,
$K_{S}^{0}\pi^{-}\pi^{+}\pi^{0}$ and $K^{-}K^{+}$.
$\overline{D}^{*0}$ mesons are reconstructed by combining the 
$\overline{D}^{0}$ candidates with a pion or a photon.
The invariant mass of $\overline{D}^{*0}$ candidates is required to be within a 
$\pm 3~\mbox{MeV}/c^{2}$ (for $\overline{D}^{*0}\pi^{0}$) and 
$\pm 10~\mbox{MeV}/c^{2}$ (for $\overline{D}^{*0}\gamma$) intervals around the nominal
$\overline{D}^{*0}$ mass. 
$D_{s}^{+}$ candidates are reconstructed in the decay modes
$D_{s}^{+}\rightarrow K_{S}^{0}K^{+}$ and $K^{+}K^{-}\pi^{+}$.
The invariant mass of the $D_{s}^{+}$ candidates is required to be within $\pm 15~\mbox{MeV}/c^{2}$ interval around the nominal $D_{s}^{+}$ mass.
$D_{s}^{*+}$ candidates are defined as $D_{s}^{+}\gamma$ combinations where the
$D_{s}^{+}\gamma$ invariant mass lies in the interval  $\pm 15~\mbox{MeV}/c^{2}$ around
the nominal $D_{s}^{*+}$ mass.
Charged $B$ pair events are produced from $\Upsilon(4S)$ resonance
$(\sqrt{s}\sim10.58~\textrm{GeV})$, where the $B^{+}$ or $B^{-}$ is produced with specific momentum and energy. 
Selection of the fully reconstructed $B$ candidates
is made according to the values of two variables :
the beam-constrained mass 
$M_{\rm bc}\equiv\sqrt{E_{\rm beam}^{2} - p_{B}^{2}}$
and the energy difference $\Delta E\equiv E_{B} - E_{beam}$.
Here, $E_{B}$ and $p_{B}$ are the reconstructed energy and momentum
of the fully reconstructed $B$ candidate in the center-of-mass (CM) system,
and $E_{\rm beam}$ is the beam energy in the CM frame.
The signal region for tagging $B$ candidates is defined as 
$M_{\rm bc}>5.27~\mbox{GeV}/c^{2}$ and $-80~\mbox{MeV}<\Delta E< 60~\mbox{MeV}$.

\begin{figure}
\centerline{
\epsfxsize=7cm \epsfbox{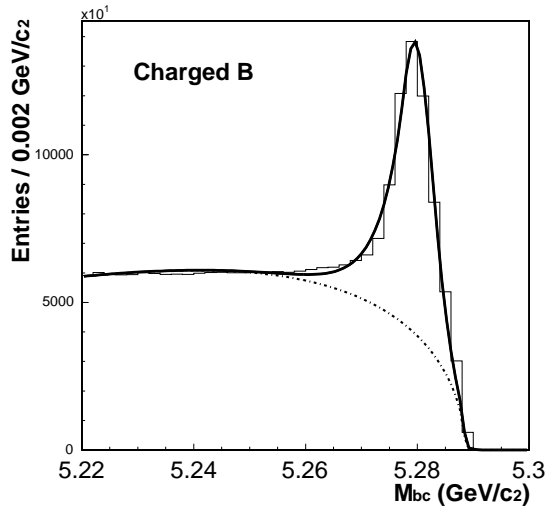} 
}
\caption{Distribution of the beam energy constrained mass in data for
	fully reconstructed $B$ mesons (histogram). The solid curve shows the 
	result of the fit and the dotted one is the background component.}
    \label{mbc_fit_ver3_dE008}
\end{figure}

The $M_{\rm bc}$ distribution of reconstructed $B$ candidates is fit with
the sum of an Argus function \cite{Albrecht:1986nr} 
and a Crystal Ball function \cite{Bloom:1983pc}.
The Argus function models the continuum and combinatorial background whereas
the Crystal Ball function models the signal component, which peaks at the
$B$ mass. The purity is defined as $S/(S+B)$, where $S~(B)$ is the number of 
signal (background) events for $M_{\rm bc} > 5.27~\textrm{GeV}/c^{2}$, 
as determined from a fit. 
Fig. \ref{mbc_fit_ver3_dE008} shows the $M_{\rm bc}$ distribution 
for all $B_{\rm rec}$ candidates in our data set in the signal $\Delta E$ region.
In this sample, there are cross-feed effects between charged and neutral $B$.
From the Monte Carlo simulation, we estimate the fraction of $B^{0}~(B^{+})$
events in the reconstruction of $B^{+}~(B^{0})$ to be $0.095~(0.090)$.
Then, we obtain $N_{B^{+}B^{-}} = (4.00\pm 0.24)\times 10^{5}$ and the purity of $0.55$, 
where the uncertainty on $N_{B^{+}B^{-}}$ is dominated by systematic errors.

In the events where a $B_{\rm rec}$ is reconstructed, we search for decays
into a $\tau$ plus a neutrino and a $K$ plus two neutrinos. 
Candidate events are required to have one or three signal-side charged
track(s) with the total charge being opposite to that of the reconstructed $B$.
The $\tau$ lepton is identified in the 
following decay channels: $\tau^{-}\rightarrow\mu^{-}\nu\bar{\nu}$,
$\tau^{-}\rightarrow e^{-}\nu\bar{\nu}$, 
$\tau^{-}\rightarrow\pi^{-}\nu$,
$\tau^{-}\rightarrow\pi^{-}\pi^{0}\nu$, and 
$\tau^{-}\rightarrow\pi^{-}\pi^{+}\pi^{-}\nu$.

We require the charged particles to be identified as leptons, pions or kaons.
The event is required to have zero net charge and $E_{\rm ECL}$ less than 
$0.3~\textrm{GeV}$ 
where $E_{\rm ECL}$ is the remaining energy calculated by adding the energy 
of the photons that are not associated with either the $B_{\rm rec}$ or 
the $\pi^{0}$ candidate from $\tau^{-}\rightarrow \pi^{-}\pi^{0}\nu$ decay.
For all modes except $\tau^{-}\rightarrow\pi^{-}\pi^{0}\nu$ mode
we reject events with $\pi^{0}$ mesons in the recoil against $B_{\rm rec}$.

We place the following requirements on the momentum
of the track(s) in the CM,
$p_{\pi^{-}} > 0.8~\textrm{GeV}/c$ for $\tau^{-}\rightarrow\pi^{-}\nu$,
$p_{\pi^{-}\pi^{0}} > 1.2~\textrm{GeV}/c$ for $\tau^{-}\rightarrow\pi^{-}\pi^{0}\nu$,
$p_{\pi^{-}\pi^{+}\pi^{-}} > 1.4~\textrm{GeV}/c$ for
$\tau^{-}\rightarrow\pi^{-}\pi^{+}\pi^{-}\nu$, and
$p_{K^{-}} > 1.2~\textrm{GeV}/c$ for $B^{-}\rightarrow K^{-}\nu\bar{\nu}$.
The event is required to have the total missing momentum of the event to be 
greater than $0.2~\textrm{GeV}/c$
for all modes except leptonic decay modes
 and the direction of missing momentum to be $-0.86 < \cos\theta_{\rm miss}^{*} < 0.95$
in the CM frame.
Further requirements are made on the invariant mass of two or three pions
$|M_{\pi\pi}-M_{\rho}| < 0.15~\mbox{GeV}/c^{2}$ and 
$|M_{\pi\pi\pi}-M_{a_{1}^{+}}| < 0.2~\mbox{GeV}/c^{2}$.
The selection efficiencies for each decay mode we consider in this
analysis are determined from a large sample of GEANT-based Monte Carlo simulations
\cite{GEANT}
for $B^{-}\rightarrow\tau^{-}\bar{\nu}$  and $B^{-}\rightarrow K^{-}\nu\bar{\nu}$ events
generated by EvtGen decay package\cite{EvtGen}.
We compute the efficiency as the ratio of the number of events surviving 
each of our selections over the number of fully reconstructed $B^{\pm}$.

The most powerful variable for separating signal and background is the remaining
energy $E_{\rm ECL}$. 
We use different energy cuts for neutral clusters contributing to the $E_{\rm ECL}$
for barrel part and end-cap parts since the effect of beam
background is severe in the end-caps.
For signal events the neutral clusters contributing to the $E_{\rm ECL}$ can only come 
from beam background, therefore the signal events peaks at low $E_{\rm ECL}$ and
the background events, which contain additional sources of neutral clusters, are
distributed toward higher $E_{\rm ECL}$ values.

The $E_{\rm ECL} < 0.3~\mbox{GeV}$ region is defined as the signal region and
the $0.45 < E_{\rm ECL} < 1.5~\mbox{GeV}$ region is defined as the sideband region.
The $E_{\rm ECL}$ shape in the MC distribution is used to extrapolate the sideband data
to the signal region.
The number of MC events in signal region and sideband are counted and their
ratio ($r_{MC}$) is obtained.
Using the number of data in the sideband and the ratio $r_{MC}$, the number of
expected background events in the signal region is estimated.
The background estimation for the different subdecay modes from the $E_{\rm ECL}$
sideband extrapolation is shown in Table \ref{tab:bg_sg_extrapolate}.
The numbers of events in sideband region agrees well between MC and data. 
To obtain the background expected from the MC simulation, $B\overline{B}$ and 
$e^+e^-\rightarrow u\bar{u},~d\bar{d},~s\bar{s},~c\bar{c}$ events are scaled to 
equivalent luminosity in data.

\begin{table}
 \begin{center}
  \renewcommand{\baselinestretch}{1.3}
   \begin{normalsize}
    \begin{tabular}{|c|c|cccc|c|} \hline
 \multicolumn{2}{|c|}{Decay Mode}  &~$r_{MC}$~  &~Sideband Data &~Sideband MC
 &~MC signal region &~Expected BG\\
\hline\hline
\lww{$B^{-}\rightarrow\tau^{-}\bar{\nu}$} 
&$\tau^{-}\rightarrow\mu^{-}\nu\bar{\nu}$ &$0.17$    &$70$  &$63.9\pm 7.2$  &$10.7\pm 2.8$  &$11.8 \pm 3.6$\\
&$\tau^{-}\rightarrow e^{-}\nu\bar{\nu}$  &$0.14$    &$67$  &$62.3\pm 8.1$   &$8.9\pm 2.6$   &$9.5 \pm 3.2$\\
&$\tau^{-}\rightarrow\pi^{-}\nu$          &$0.07$    &$47$  &$48.4\pm 7.7$   &$3.6\pm 1.6$   &$3.5 \pm 1.7$\\
&$\tau^{-}\rightarrow\pi^{-}\pi^{0}\nu$   &$0.23$    &$13$  &$19.2\pm 4.0$   &$4.4\pm 2.1$   &$3.0 \pm 1.8$\\
&$\tau^{-}\rightarrow\pi^{-}\pi^{+}\pi^{-}\nu$ &$0.23$    &$16$  &$23.0\pm 7.2$   &$5.2\pm 2.5$   &$3.6 \pm 2.2$\\
\hline
\multicolumn{2}{|c|}{$B^{-}\rightarrow K^{-}\nu\bar{\nu}$} &$0.15$    &$17$  &$18.9\pm 4.7$   &$2.9\pm 1.4$   &$2.6 \pm 1.6$\\
\hline
    \end{tabular}
    \caption{Expected background in the signal region for the different selection modes.}
   \label{tab:bg_sg_extrapolate}
  \end{normalsize}
 \end{center}
\end{table}

\begin{figure}
\centerline{
\epsfxsize=8cm \epsfbox{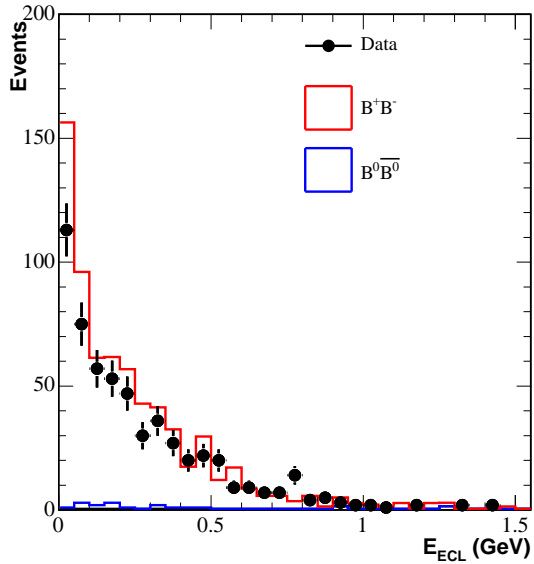} 
}
\caption{The $E_{\rm ECL}$ distribution for double-tagged events, plotted for simulation and
  data. Continuum background is negligible.}
    \label{ecl_semilepton}
\end{figure}

The double tag events, for which one of the $B$ mesons is fully reconstructed and
the other $B$ meson is reconstructed in the set of decay modes
$B^{-} \rightarrow D^{0}\ell^{-}\bar{\nu}$, where $\ell$ is a muon or an electron
and the $D^{0}$ is reconstructed in two modes : 
$D^{0}  \rightarrow K^{+}\pi^{-}~\& ~K^{+}\pi^{-}\pi^{+}\pi^{-}$,
are used as a control sample to validate the $E_{\rm ECL}$ simulation.
The sources affecting the $E_{\rm ECL}$ in double-tagged events are
similar to those affecting the $E_{\rm ECL}$ distribution in 
the signal MC simulation.
The agreement of the $E_{\rm ECL}$ distribution between data and MC simulation for the
double-tagged sample, in Fig. \ref{ecl_semilepton}, is used as a validation of the
$E_{\rm ECL}$ simulation in the signal MC.

The main sources of uncertainty we consider in the determination of the
${\cal B}(B^{-}\rightarrow\tau^{-}\overline{\nu})$ and
${\cal B}(B^{-}\rightarrow K^{-}\nu^{-}\overline{\nu})$ are 
the number of $B^{+}B^{-}$ events with one reconstructed $B$,
the determination of the signal efficiency and
the determination of the number of expected background events.
The number of $B^{+}B^{-}$ events is determined as the area 
of the Crystal Ball function fitted to the $M_{\rm bc}$ distribution. 
Using a Gaussian function as an alternative fitting function,
we obtain a relative change in the number of events and
this difference is assigned as the systematic uncertainty on  
the number of $B^{+}B^{-}$ events.
The main contribution to the systematic uncertainties in the determination 
of the efficiencies comes from uncertainty on tracking efficiency, 
Monte Carlo statistics and particle identification.
The uncertainty in the expected background comes from Monte Carlo statistics
and statistics of sideband data events.
Estimates of systematic uncertainties are summarized in 
Table \ref{tab:total_systematics}. 

\begin{table}
 \begin{center}
  \renewcommand{\baselinestretch}{1.3}
   \begin{normalsize}
    \begin{tabular}{|c|cccccc|} \hline
       Source   &\multicolumn{6}{c|}{Relative uncertainty $(\%)$}\\\hline\hline   
  &$\mu^{-}\nu\bar{\nu}$ &$ e^{-}\nu\bar{\nu}$
  &$\pi^{-}\nu$          &$\pi^{-}\pi^{0}\nu$  
  &$\pi^{+}\pi^{-}\pi^{+}\nu$ &$K^{-}\nu\bar{\nu}$  \\  \hline\hline
 Number of $B^{+}B^{-}$  &\multicolumn{6}{c|}{$6.0$}           \\\hline\hline
 tracking     &$2.0$   &$2.0$   &$2.0$   &$2.0$   &$6.0$   &$2.0$ \\
 $\tau$ decay BR  &$0.3$   &$0.3$   &$1.0$   &$0.6$   &$1.1$   &\\
 MC statistics    &$0.6$   &$0.6$   &$0.7$   &$1.0$   &$2.0$   &$4.1$\\
 Lepton identification        &$2.2$   &$2.2$   &        &        &  &   \\
 $\pi^{0}$ identification    &        &        &        &$2.6$   & &    \\
 $\pi^{\pm}$ identification  &        &        &$1.7$   &$1.7$   &$5.1$ &\\
 Total Efficiency Uncertainty &$2.9$ &$2.9$ &$3.0$ &$4.0$ &$8.8$ &$5.0$ \\\hline\hline
 MC statistics    &$28.2$ &$31.8$ &$47.5$ &$52.0$ &$57.1$  &$55.8$\\
 Data in sideband &$12.0$ &$12.2$ &$14.6$ &$27.2$ &$25.0$  &$24.3$\\
 Total Background Uncertainty &$30.6$ &$34.1$ &$49.7$ &$58.7$ &$62.3$ &$60.9$\\\hline
   \end{tabular}
   \caption{Systematic uncertainties for the number of $B^{+}B^{-}$ events
	 with one reconstructed $B$, the determination of the efficiency and 
         the determination of the number of expected background events for
	the different decay channels. }
   \label{tab:total_systematics}
  \end{normalsize}
 \end{center}
\end{table}

\begin{figure}
\centerline{
  \epsfxsize=18cm \epsfbox{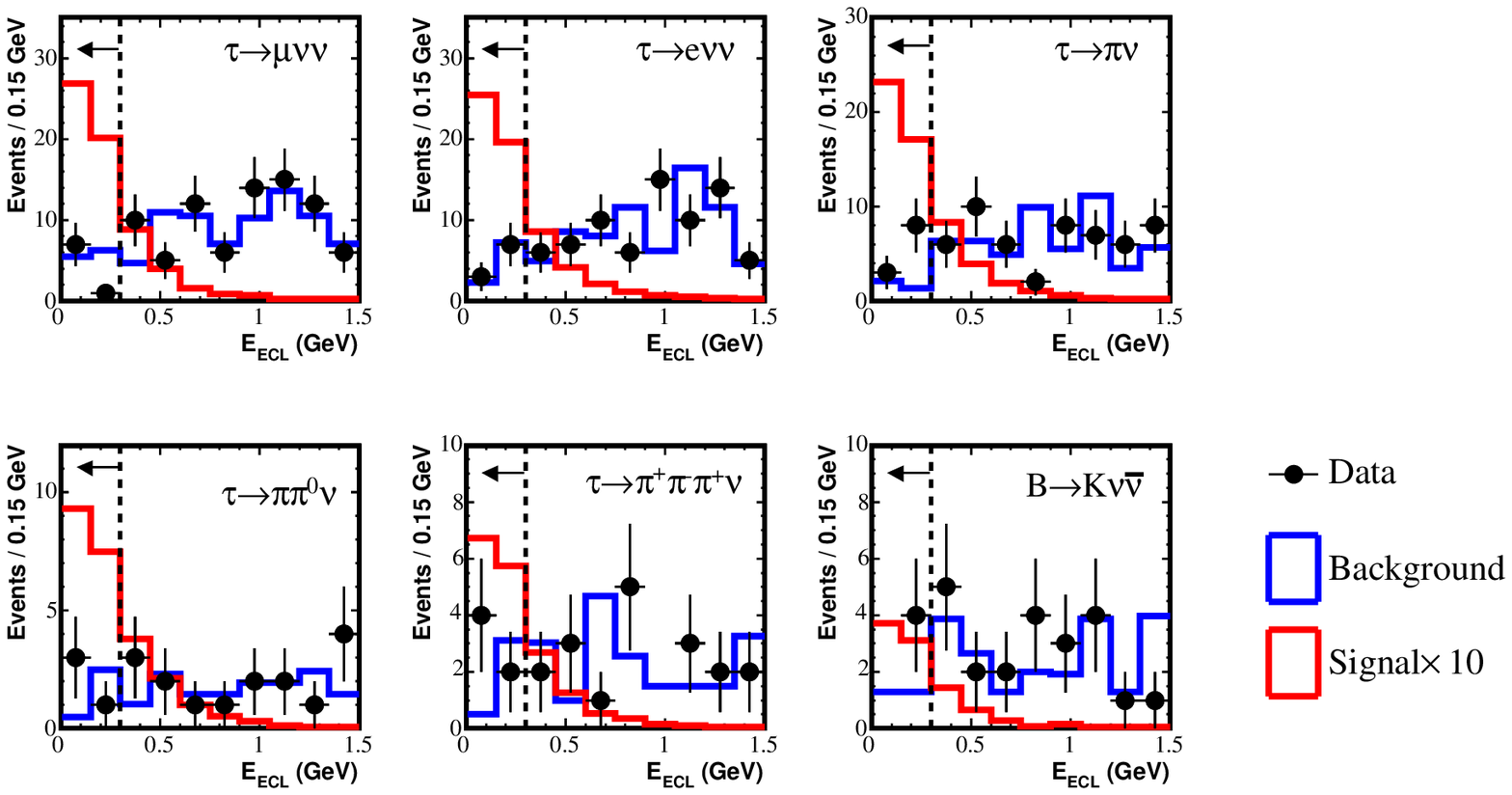} 
}
\caption{$E_{\rm ECL}$ distributions in the data after
	all selection requirements except the one on $E_{\rm ECL}$ 
	have been applied. The red histograms represent the distributions for
        Monte Carlo simulated signal events scaled to ${\cal B}(\Btaunu)=10^{-4}$ and ${\cal B}(\BKnunu)=4\times10^{-6}$.}
    \label{ecl_opened}
\end{figure}

After finalizing the signal selection criteria, the signal region 
($E_{\rm ECL} < 0.3~\mbox{GeV}$) in the on-resonance data is examined.
Table \ref{tab:observed_events} lists the number of observed events in data in the
signal region, together with the expected number of signal and background events
in the signal region.
Fig. \ref{ecl_opened} shows the $E_{\rm ECL}$ distributions in the data after
all selection requirements except the one on $E_{\rm ECL}$ have been applied
compared with the expected background. Each distribution refers to 
a different mode.

\begin{table}
 \begin{center}
  \renewcommand{\baselinestretch}{1.3}
   \begin{normalsize}
    \begin{tabular}{c|c|cccc} \hline
 \multicolumn{2}{c|}{\lw{Decay Mode}}  
  &Signal     &Signal     &Background &Observed \\
 \multicolumn{2}{c|}{} 
  &Efficiency$(\%)$    &Expected   &Expected &Events   \\\hline \hline 
\lww{$B^{-}\rightarrow\tau^{-}\bar{\nu}$} 
&$\tau^{-}\rightarrow\mu^{-}\nu\bar{\nu}$ &$ 9.8\pm 0.1$ &$ 3.9\pm 0.1$ &$11.8\pm 3.6$ &$8$\\
&$\tau^{-}\rightarrow e^{-}\nu\bar{\nu}$  &$ 9.4\pm 0.1$ &$ 3.8\pm 0.1$ &$9.5\pm 3.2$  &$10$\\
&$\tau^{-}\rightarrow\pi^{-}\nu$          &$ 8.4\pm 0.1$ &$ 3.4\pm 0.1$ &$3.5\pm 1.7$  &$11$\\
&$\tau^{-}\rightarrow\pi^{-}\pi^{0}\nu$   &$ 3.5\pm 0.1$ &$ 1.4\pm 0.1$ &$3.0\pm 1.8$  &$4$\\
&$\tau^{-}\rightarrow\pi^{-}\pi^{+}\pi^{-}\nu$ &$ 2.6\pm 0.1$ &$ 1.0\pm 0.1$ &$3.6\pm 2.2$  &$6$\\
\hline
\multicolumn{2}{c|}{Total} &$ 33.7\pm 1.4$ &$ 13.5\pm 0.2$ &$31.4\pm 5.9$  &$39$\\
\hline\hline
\multicolumn{2}{c|}{$B^{-}\rightarrow K^{-}\nu\bar{\nu}$} &$ 42.8\pm 1.8$ &$ 0.70\pm 0.03$ &$2.6\pm 1.6$  &$4$\\
\hline
    \end{tabular}
    \caption{Number of observed data events in the signal region, together with
    number of expected signal and background events.
    Errors in the background expectation is both statistical and 
    systematic errors. The numbers of expected signal are obtained by assuming that
    ${\cal B}(\Btaunu)=10^{-4}$ and ${\cal B}(\BKnunu)=4\times10^{-6}$. }
   \label{tab:observed_events}
  \end{normalsize}
 \end{center}
\end{table}

\begin{figure}
\centerline{
\epsfxsize=8cm \epsfbox{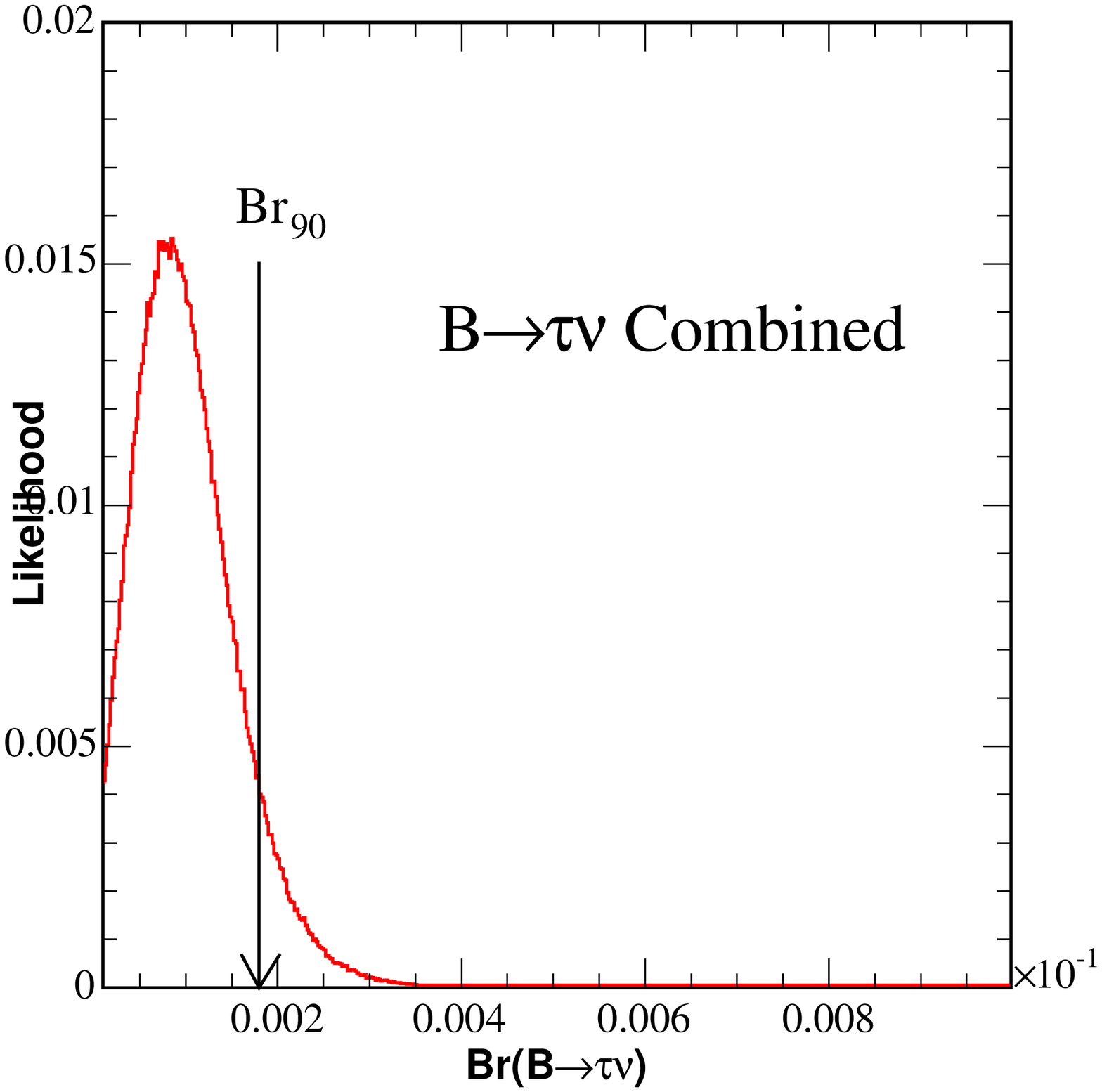}
\epsfxsize=8cm \epsfbox{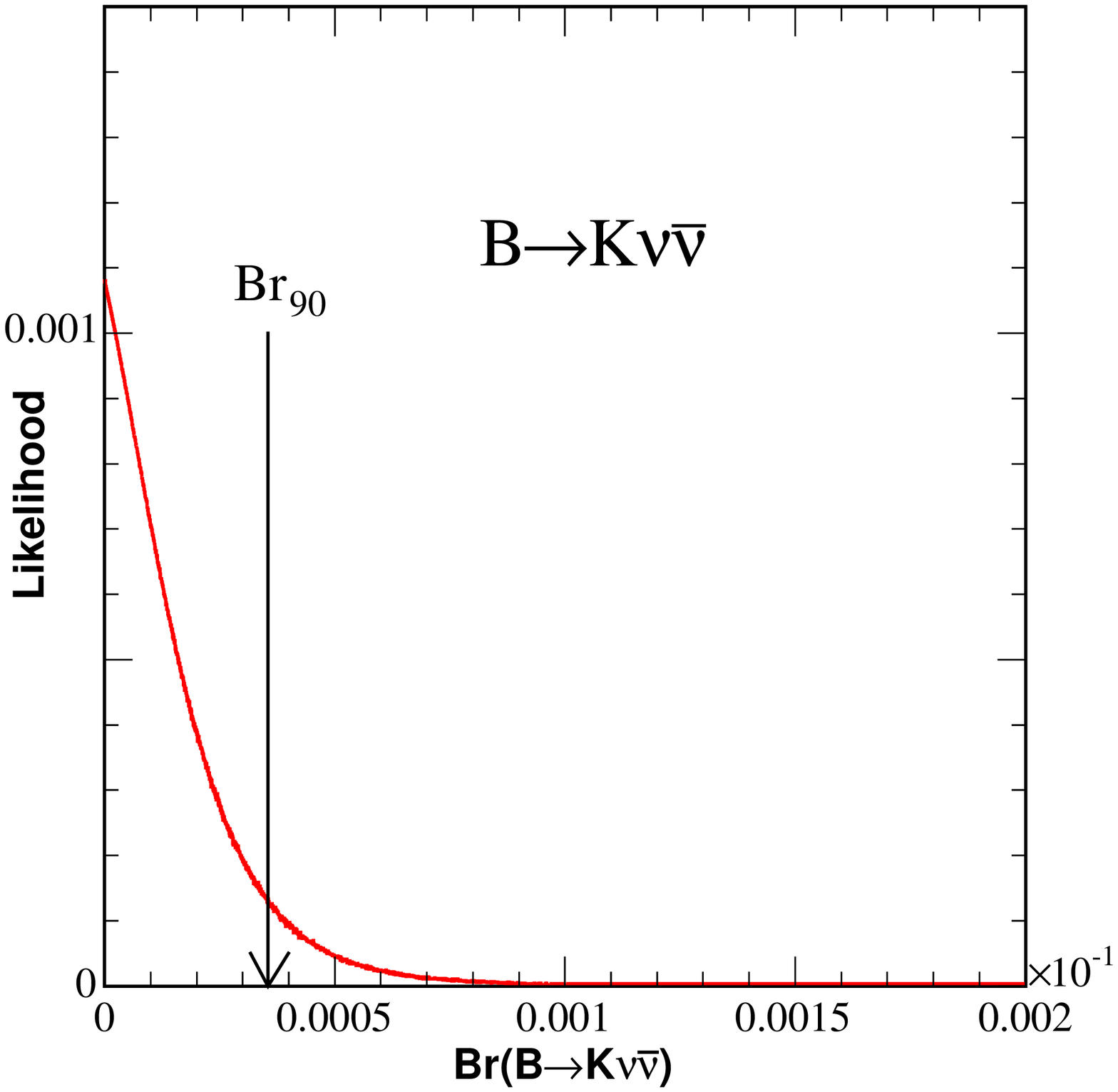}
}
    \caption{Likelihood functions for the fit to data after smearing.}
    \label{taunu_like_landau}
\end{figure}

Since we do not observe significant excess over the expected background, we set
upper limits.
To extract the upper limits on the branching fraction for
${\cal B}(B^{-}\rightarrow\tau^{-}\overline{\nu})$ and
${\cal B}(B^{-}\rightarrow K^{-}\nu\overline{\nu})$, we fit the observed
$E_{\rm ECL}$ distributions to the expected background and signal, using maximum likelihood
method. The likelihood function ${\cal L}$ is defined as
$$
{\cal L} = \frac{1}{\sqrt{2\pi}\sigma_{b}}e^{-(n_{b}-N_{b})^{2}/2\sigma_{b}}
\cdot \frac{e^{-(n_{s}+n_{b})}(n_{s}+n_{b})^{N}}{N!}\prod_{i=1}^{N}\frac{n_{s}f_{b}(i)+n_{b}f_{s}(i)}{n_{s}+n_{b}}
$$
where $n_{b}$ and $n_{s}$ represent the number of background and signal events, 
respectively, $N$ is the number of observed events, $N_{b}$ is the calculated number of
background events, and $\sigma_{b}$ is the calculated background uncertainty.
The variable $f_{s}$ is the normalized signal $E_{\rm ECL}$ distribution and $f_{b}$
is the normalized background $E_{\rm ECL}$ distribution.
The negative log likelihood function is minimized using MINUIT \cite{James:1975dr}
with respect to
$n_{b}$ for each $n_{s}~(=\varepsilon_{i}\cdot N_{B^{+}B^{-}}\cdot {\cal B})$.
The $90\%$ C. L. upper limit on the branching fraction ${\cal B}$ is calculated by
\begin{equation}
 \label{eq:alpha_integrate}
 0.9 = \frac{\int_{0}^{{\cal B}_{90}}{\cal L}({\cal B})d{\cal B}}
             {\int_{0}^{\infty}{\cal L}({\cal B})d{\cal B}}
\end{equation}
For $\Btaunu$, we calculate the likelihood function for each different decay 
mode (${\cal L}_{i}({\cal B})$).
Total likelihood function is defined by
\begin{equation}
 \label{eq:L_alpha}
 {\cal L}({\cal B})  =  \prod_{i=1}^{n_{ch}} {\cal L}_{i}({\cal B})
\end{equation}
where $n_{ch}$ is the number of decay modes for $\Btaunu$.
The full systematic uncertainty must be incorporated into the likelihood
function. We convolve the systematic uncertainty into the likelihood function,
${\cal L}({\cal B})$, by replacing each point of ${\cal L}({\cal B})$ by a Gaussian distribution centered at
that point with width $\Delta{\cal B}$ which is determined from systematic
uncertainty study.
To get the value of a particular point of the smeared likelihood function,
we integrate all the contributions from the Gaussian replaced points of
the unsmeared likelihood function.
To combine likelihood functions of 5 decay modes for $\Btaunu$, we simply 
multiply the likelihood functions to produce the combined likelihood.
The only complication arises in that
there are common sources of systematic uncertainty and therefore correlated
uncertainties between the samples.
We throw different normal Gaussian random number for correlated
and uncorrelated systematics for each systematics sources.
In this way, we can smear the correlated systematics in the same direction.
We obtain the final smeared likelihood function by multiplying smeared
likelihood functions for $\Btaunu$.
For $\BKnunu$ decay, we smear the likelihood function by the total systematic
uncertainty.
Fig. \ref{taunu_like_landau} shows the likelihood functions for the fit to data
after smearing for $\Btaunu$ (left) and $\BKnunu$ (right).
We obtain upper limits on the branching fraction 
at the 90\% (C.L.) of
\begin{equation}
{\cal B}(B^{-}\rightarrow\tau^{-}\overline{\nu}) < 1.8\times 10^{-4}
\end{equation}
\begin{equation}
{\cal B}(B^{-}\rightarrow K^{-}\nu\overline{\nu}) < 3.6\times 10^{-5}.
\end{equation}

\begin{figure}
\centerline{
\epsfxsize=7cm \epsfbox{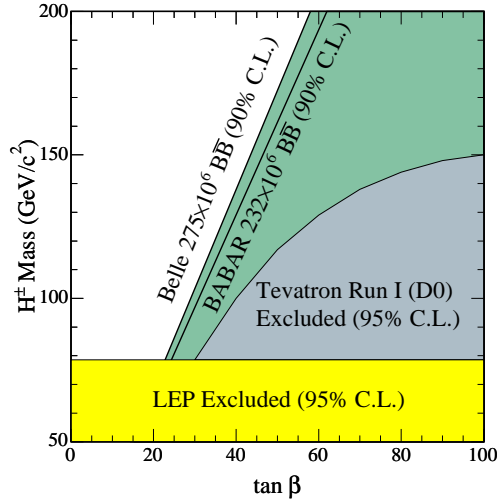}
}
    \caption{The $90\%$ C.L. exclusion boundaries in the 
	$[M_{H^{+}}, \tan\beta]$ plane obtained from the observed upper 
	limit on ${\cal B}(B^{-}\rightarrow \tau^{-}\nu)$.}
    \label{Mh_tanb_LP05}
\end{figure}

In the extension of the Standard Model, one expects significant modification to
the $\Btaunu$ decay branching fraction.
In the two-Higgs doublet model, the decay can occur via a charged 
Higgs particle. The $\Btaunu$ branching fraction is given as
\begin{equation}
{\cal B}(\Btaunu)
 = {\cal B}(\Btaunu)_{\rm SM} \times r_{H},
\end{equation}
where $r_{H}$ is defined as
\begin{equation}
r_{H} = \left(1-\frac{\tan^{2}\beta}{m_{H}^{2}}m_{B}^{2}\right)^{2},
\label{eq:r_H}
\end{equation}
$m_{H}$ is the charged Higgs mass and 
$\tan\beta$ is the ratio of vacuum expectation values of two Higgs 
doublets \cite{Hou:1992sy}. ${\cal B}(\Btaunu)_{\rm SM}$ represents SM contribution
given by Eq.(\ref{eq:BR_B_taunu}).
Once we get an upper limit on ${\cal B}(\Btaunu)$,
we can give a constraint on $\tan\beta$ and $m_{H}$.
Fig. \ref{Mh_tanb_LP05} shows the $90\%$ C.L. exclusion 
boundaries in the $[m_{H}, \tan\beta]$ plane obtained 
with $m_{B}=5279~\textrm{MeV}/c^{2}$ and 
${\cal B}(\Btaunu)_{\rm SM} = 0.93\times 10^{-4}$ from the CKMfitter prediction
compared with other experimental searches
at LEP\cite{Bock:2000gk}, at the Tevatron\cite{Abazov:2001md}, and at BABAR.

In conclusion, we have performed a search for 
the $B^{-}\rightarrow\tau^{-}\overline{\nu}$ 
and $B^{-}\rightarrow K^{-}\nu\overline{\nu}$ decays in a fully reconstructed $B$
sample. 
Upper limits have been set :
\begin{equation}
{\cal B}(B^{-}\rightarrow\tau^{-}\overline{\nu}) < 1.8\times 10^{-4}~~(90\%~\textrm{C.L.})
\end{equation}
\begin{equation}
{\cal B}(B^{-}\rightarrow K^{-}\nu\overline{\nu}) < 3.6\times 10^{-5}~~(90\%~\textrm{C.L.})
\end{equation}
which are the most stringent upper limits on these processes to date.

We thank the KEKB group for the excellent operation of the
accelerator, the KEK cryogenics group for the efficient
operation of the solenoid, and the KEK computer group and
the National Institute of Informatics for valuable computing
and Super-SINET network support. We acknowledge support from
the Ministry of Education, Culture, Sports, Science, and
Technology of Japan and the Japan Society for the Promotion
of Science; the Australian Research Council and the
Australian Department of Education, Science and Training;
the National Science Foundation of China under contract
No.~10175071; the Department of Science and Technology of
India; the BK21 program of the Ministry of Education of
Korea and the CHEP SRC program of the Korea Science and
Engineering Foundation; the Polish State Committee for
Scientific Research under contract No.~2P03B 01324; the
Ministry of Science and Technology of the Russian
Federation; the Ministry of Higher Education, 
Science and Technology of the Republic of Slovenia;  
the Swiss National Science Foundation; the National Science Council and
the Ministry of Education of Taiwan; and the U.S.\
Department of Energy.


\end{document}

%% file: author-conf2005.tex
\affiliation{Aomori University, Aomori}
\affiliation{Budker Institute of Nuclear Physics, Novosibirsk}
\affiliation{Chiba University, Chiba}
\affiliation{Chonnam National University, Kwangju}
\affiliation{University of Cincinnati, Cincinnati, Ohio 45221}
\affiliation{University of Frankfurt, Frankfurt}
\affiliation{Gyeongsang National University, Chinju}
\affiliation{University of Hawaii, Honolulu, Hawaii 96822}
\affiliation{High Energy Accelerator Research Organization (KEK), Tsukuba}
\affiliation{Hiroshima Institute of Technology, Hiroshima}
\affiliation{Institute of High Energy Physics, Chinese Academy of Sciences, Beijing}
\affiliation{Institute of High Energy Physics, Vienna}
\affiliation{Institute for Theoretical and Experimental Physics, Moscow}
\affiliation{J. Stefan Institute, Ljubljana}
\affiliation{Kanagawa University, Yokohama}
\affiliation{Korea University, Seoul}
\affiliation{Kyoto University, Kyoto}
\affiliation{Kyungpook National University, Taegu}
\affiliation{Swiss Federal Institute of Technology of Lausanne, EPFL, Lausanne}
\affiliation{University of Ljubljana, Ljubljana}
\affiliation{University of Maribor, Maribor}
\affiliation{University of Melbourne, Victoria}
\affiliation{Nagoya University, Nagoya}
\affiliation{Nara Women's University, Nara}
\affiliation{National Central University, Chung-li}
\affiliation{National Kaohsiung Normal University, Kaohsiung}
\affiliation{National United University, Miao Li}
\affiliation{Department of Physics, National Taiwan University, Taipei}
\affiliation{H. Niewodniczanski Institute of Nuclear Physics, Krakow}
\affiliation{Nippon Dental University, Niigata}
\affiliation{Niigata University, Niigata}
\affiliation{Nova Gorica Polytechnic, Nova Gorica}
\affiliation{Osaka City University, Osaka}
\affiliation{Osaka University, Osaka}
\affiliation{Panjab University, Chandigarh}
\affiliation{Peking University, Beijing}
\affiliation{Princeton University, Princeton, New Jersey 08544}
\affiliation{RIKEN BNL Research Center, Upton, New York 11973}
\affiliation{Saga University, Saga}
\affiliation{University of Science and Technology of China, Hefei}
\affiliation{Seoul National University, Seoul}
\affiliation{Shinshu University, Nagano}
\affiliation{Sungkyunkwan University, Suwon}
\affiliation{University of Sydney, Sydney NSW}
\affiliation{Tata Institute of Fundamental Research, Bombay}
\affiliation{Toho University, Funabashi}
\affiliation{Tohoku Gakuin University, Tagajo}
\affiliation{Tohoku University, Sendai}
\affiliation{Department of Physics, University of Tokyo, Tokyo}
\affiliation{Tokyo Institute of Technology, Tokyo}
\affiliation{Tokyo Metropolitan University, Tokyo}
\affiliation{Tokyo University of Agriculture and Technology, Tokyo}
\affiliation{Toyama National College of Maritime Technology, Toyama}
\affiliation{University of Tsukuba, Tsukuba}
\affiliation{Utkal University, Bhubaneswer}
\affiliation{Virginia Polytechnic Institute and State University, Blacksburg, Virginia 24061}
\affiliation{Yonsei University, Seoul}
  \author{K.~Abe}\affiliation{High Energy Accelerator Research Organization (KEK), Tsukuba} 
  \author{K.~Abe}\affiliation{Tohoku Gakuin University, Tagajo} 
  \author{I.~Adachi}\affiliation{High Energy Accelerator Research Organization (KEK), Tsukuba} 
  \author{H.~Aihara}\affiliation{Department of Physics, University of Tokyo, Tokyo} 
  \author{K.~Aoki}\affiliation{Nagoya University, Nagoya} 
  \author{K.~Arinstein}\affiliation{Budker Institute of Nuclear Physics, Novosibirsk} 
  \author{Y.~Asano}\affiliation{University of Tsukuba, Tsukuba} 
  \author{T.~Aso}\affiliation{Toyama National College of Maritime Technology, Toyama} 
  \author{V.~Aulchenko}\affiliation{Budker Institute of Nuclear Physics, Novosibirsk} 
  \author{T.~Aushev}\affiliation{Institute for Theoretical and Experimental Physics, Moscow} 
% ...
\collaboration{The Belle Collaboration}

%% file: Btaunu_LP05_new.bbl
\begin{thebibliography}{99}

\bibitem{Charles:2004jd}
  J.~Charles {\it et al.}  (CKMfitter Group),
  Eur.\ Phys.\ J.\ C {\bf 41}, 1 (2005) and the updated results presented at
  CKM2005 workshop.


\bibitem{Aubert:2004kz}
B.~Aubert  (BABAR Collaboration),
arXiv:hep-ex/0407038.


\bibitem{Grossman:1995gt}
  Y.~Grossman, Z.~Ligeti and E.~Nardi,
  Nucl.\ Phys.\ B {\bf 465}, 369 (1996)
  [Erratum-ibid.\ B {\bf 480}, 753 (1996)]
  [arXiv:hep-ph/9510378].


\bibitem{Bird:2004ts}
  C.~Bird, P.~Jackson, R.~Kowalewski and M.~Pospelov,
  Phys.\ Rev.\ Lett.\  {\bf 93}, 201803 (2004)
  [arXiv:hep-ph/0401195].


\bibitem{Faessler:2002ut}
  A.~Faessler, T.~Gutsche, M.~A.~Ivanov, J.~G.~Korner and V.~E.~Lyubovitskij,
  Eur.\ Phys.\ J.\ directC {\bf 4}, 18 (2002)
  [arXiv:hep-ph/0205287].


\bibitem{Buchalla:2000sk}
  G.~Buchalla, G.~Hiller and G.~Isidori,
  Phys.\ Rev.\ D {\bf 63}, 014015 (2001)
  [arXiv:hep-ph/0006136].


\bibitem{Aubert:2004ws}
  B.~Aubert {\it et al.}  (BABAR Collaboration),
  Phys.\ Rev.\ Lett.\  {\bf 94}, 101801 (2005)
  [arXiv:hep-ex/0411061].


\bibitem{Kurokawa:2003}
S.~Kurokawa and E.~Kikutani,
Nucl.\ Instrum.\ Methods\ Phys.\ Res.,\ Sect.\ A {\bf 499}, 1 (2003)



\bibitem{belle_detector:2003}
Belle Collaboration, A.~Abashian {\it et al.}
Nucl.\ Instrum.\ Methods\ Phys.\ Res.,\ Sect.\ A {\bf 479}, 117 (2002)



\bibitem{Albrecht:1986nr}
H.~Albrecht {\it et al.}  (ARGUS Collaboration),
Phys.\ Lett.\ B {\bf 185}, 218 (1987).



\bibitem{Bloom:1983pc}
E.~D.~Bloom and C.~Peck,
Ann.\ Rev.\ Nucl.\ Part.\ Sci.\  {\bf 33}, 143 (1983).


\bibitem{GEANT}
R.~Brun {\it et al.},
GEANT3.21,~CERN Report DD/EE/84-1 (1984).


\bibitem{EvtGen}
See the EvtGen package home page,
http://www.slac.stanford.edu/\~{ }lange/EvtGen/.


\bibitem{James:1975dr}
  F.~James and M.~Roos,
  Comput.\ Phys.\ Commun.\  {\bf 10} (1975) 343.

\bibitem{Hou:1992sy}
W.~S.~Hou,
Phys.\ Rev.\ D {\bf 48}, 2342 (1993).


\bibitem{Bock:2000gk}
P.~Bock {\it et al.}  (ALEPH, DELPHI, L3 and OPAL Collaborations),
CERN-EP-2000-055


\bibitem{Abazov:2001md}
V.~M.~Abazov {\it et al.}  (D0 Collaboration),
Phys.\ Rev.\ Lett.\  {\bf 88}, 151803 (2002)
[arXiv:hep-ex/0102039].


\end{thebibliography}
